\documentstyle[prd,aps,floats]{revtex}
\begin{document}
\draft
\preprint{SUSSEX-AST 98/6-1, astro-ph/9909008}
 
%
% Remove this and closure after abstract, plus preprint number,
% in electronic submission
%
\input epsf
\renewcommand{\topfraction}{0.8}
%\twocolumn[\hsize\textwidth\columnwidth\hsize\csname
%@twocolumnfalse\endcsname
 
\title{A quantum holographic principle from decoherence} 
\author{Sougato Bose $^ \dagger $ and Anupam Mazumdar ${^ \star}$}
\address{$\dagger $ Optics Section, Blackett Laboratory, 
Imperial College, London, SW7 2BZ, ~~U.~K. \\
$\star $ Astrophysics Group, Blackett Laboratory, Imperial College,
London, SW7 2BZ, ~~U.~K.}
\date{\today}
\maketitle
\begin{abstract}
We present a fully quantum version of the holographic principle
in terms of quantum systems, subsystems, and their interactions. 
We use the concept of enviornment induced decoherence to prove this
principle. We discuss the conditions under which the standard
(semi-classical) holographic principle is obtained from this
quantum mechanical version.
\end{abstract}
 
\pacs{PACS numbers: 98.80.Cq \hspace*{1.1cm} IMPERIAL preprint IMPERIAL-AST
99/9-5, gr-qc/yymmmnn}

\vskip2pc 
%%%%%%%%%%%%%%%%%%%%%%%%%%%%%%%%%%%%%%%%%%%%%%%%%%%%%%%%%%%%%%%%%%%%%%%%
\section{INTRODUCTION}
It was first propounded by 't Hooft \cite{thooft} that the observable degrees 
of freedom of a $3+1$ diemsnional world can as well be realised from the 
boundary of the system. This idea fits in perfectly with 
Bekenstein's result about the total entropy of a black-hole  
not exceeding a quarter of the area of its event horizon \cite{bekenstein}.
Following 't Hooft's, this 
suggests that all the information of a black hole can be 
collected from the surface of its horizon. There also exists a 
conjecture in supergravity Yang-Mills theories that the bulk information can 
be successfully stored at the boundary of an anti-de-Sitter space \cite
{maldacena}. All these results lead to the conjecture : {\it all
information about a system can be obtained from its boundary}, which is
known as the holographic principle.

However, the application of holographic principle, as it stands, becomes
restricted to situations where one can uniquely define the
boundary of a system. What could be the boundary of a purely 
quantum mechanical system?
One could imagine it to be the boundary of the classical potential which traps
the system. From such considerations, for a particle in a box, the box would be 
its boundary. However, a quantum system can always tunnel out from any finite 
classical potential. Moreover, in reality, classical ptential wells are
the results of the interaction of a quantum system with other quantum 
fields and the notion 
of a classical potential well does not even exist in an entirely quantum 
universe. This motivates us to formulate a version of the holographic principle 
entirely in terms of quantum systems and their interactions. To 
prove this version of the principle, we rely on the notion of 
environment induced decoherence (EID) \cite{zurek91}.

EID is a process used to explain the
emergence of classicality from the quantum world. When an isolated 
quantum system interacts with it's environment, its state evolves to a 
diagonal mixed states in a specific pointer basis \cite{zurek81}. 
This is also a 
mechanism for entropy production in an open quantum system. We show that
if this is the sole entropy generation mechanism 
in a system, then a quantum version of the holographic
principle can be proved. In other words, we assume that {\it  
all the entropy of a quantum 
system arises from its interactions with its environment}. This 
assumption is justified if all systems in our universe are quantum. Their
entropy cannot increase due to their own unitary evolutions. They gain 
entropy only when they interact with other systems.

We will first give an operational definition of the boundary of a quantum
system in terms of the notions of systems, sub-systems and their interactions.
Next we will show how the decoherence of the system leads to all information
about the system being stored in its boundary ( as defined by us ). After this,
we will use the stability of the states of the pointer basis in 
decoherence to justify
why the information {\it remains stored } in this way. To conclude the 
paper, we will point out the conditions required for our quantum holographic
principle to go over to the standard ( semi-classical) holographic 
principle.

%%%%%%%%%%%%%%%%%%%%%%%%%%%%%%%%%%%%%%%%%%%%%%%%%%%%%%%%%%%%%%%%%%%%%%%%%%
\section{THE QUANTUM HOLOGRAPHIC PRINCIPLE}

Let there be a quantum system $\rm S$. All quantum systems interacting with it 
together constitute another system, called the environment $\rm E$. However,
in general, all constituents of the system $\rm S$ will not interact with 
$\rm E$ directly. Therefore, we divide the system $\rm S$ into two subsystems
$\rm A$ and $\rm B$ with $\rm B$ being the subsystem which {\it directly 
interacts }
with $\rm E$. We define $\rm B$ to be the boundary of the system $\rm S$. 
Using this definition of the boundary, we can formulate the following 
holographic principle : {\it all information about $\rm S$ can be obtained
from $\rm B$}.

As we have assumed that all the entropy of the system is due to its 
interaction with its environment, we can start off the system $S$,
before any interaction with $\rm E$, in the pure state 
$|\psi \rangle _{\rm AB}$.
Let us assume that $\hat X_{\rm B}$ is the operator whose
eigenstates $|X_{i}\rangle _{\rm B}$ form the pointer basis for the 
subsystem $\rm B$. This means, that $\rm B$ interacts with  
$\rm E$ via an
interaction Hamiltonian of the type
\begin{eqnarray}
\label{red0}
H_{\rm BE} = g_{1}\hat X_{\rm B} \hat Y_{\rm E}\,,
\end{eqnarray}
where $\hat Y_{\rm E}$ is some operator in the Hilbert space of the system
$\rm E$ and $g_{1}$ is the coupling strength. If $\rm E$ is a sufficiently
large system, and if $\rm B$ has many degrees of freedom interacting with
$\rm E$, then this interaction leads to a diagonalization of the state $\rm B$
in the basis $|X_{i}\rangle $. The initial pure state of the system $\rm S$
can be written in this basis as
\begin{eqnarray}
\label{red1}
|\psi \rangle _{\rm {AB}}= \sum _{i} c_{i} |\phi_{i}\rangle_{\rm A}
|X_{i} \rangle _{\rm B}\,,
\end{eqnarray}
where the state $|\phi _{i}\rangle _{\rm A}$ need not be orthogonal,
but are assumed to be normalized. After interaction of $\rm B$ with $\rm E$
for a span of 
time more than the decoherence time scale \cite{zurek91}
this state evolves to
\begin{eqnarray}
\label{red2}
\rho_{\rm {AB}} =\sum _{i} |c_{i}|^{2} |\phi_{i} \rangle _{\rm A}
\langle \phi_{i} |_{\rm A} \otimes |X_{i}\rangle _{\rm B} \langle 
X_{i}|_{\rm B} \,.
\end{eqnarray}
At this stage the von-Neumann entropy $S^{v}(\rho_{\rm {AB}})$ of the
system $S$ is given by 
\begin{eqnarray}
\label{red3}
S^{v} (\rho_{\rm {AB}}) = -{\rm {Tr}}( \rho_{\rm {AB}} 
\ln \rho_{\rm {AB}})
= -\sum_{i} |c_{i}|^{2} \ln |c_{i}|^{2} \,,
\end{eqnarray}
The Von-Newmann entropy of the boundary $B$ is obtained from $\rho_{\rm B}
=\rm{Tr}_{\rm A}(\rho_{\rm {AB}})$ to be
\begin{eqnarray}
\label{red4}
S^{v}(\rho_{\rm B}) = -{\rm {Tr}}(\rho_{\rm B} \ln \rho_{\rm B}) =-
\sum_{i} |c_{i}|^2 \ln |c_{i}|^2  =  S^{v}(\rho_{\rm {AB}})\,.
\end{eqnarray} 
It is known that the von-Neumann entropy of a system is equal to the amount of 
information one can acquire on observing the system's state \cite{schu}.
Eq.(\ref{red4}) shows that the entire information $S^{v}
(\rho_{\rm {AB}})$ about the system $\rm S$ can be gained from just 
acquiring the information 
$S^{v}(\rho_{\rm B})$ stored on it's boundary $\rm B$.

However, how stable is this information stored in the boundary? In fact,
what about the posssibility of Eq.(\ref{red4}) loosing it's validity,
due to the interaction between subsystems $\rm A$ and $\rm B$ of the system
$\rm S$? This possibility can be prevented if $\rm B$ is a subsystem
with a large number of degrees of freedom (i.e a macroscopic system).
Then by definition of the pointer basis, the states of the basis $
|X_{i}\rangle$ are stable states. In fact, it is the {\it stability
of states in the pointer basis } \cite{zurek}, which makes them ideal
candidates for classical states. This demands that the subsystem $\rm B$
interacts with the subsystem $\rm A$ via an interaction Hamiltonian
of the type
\begin{eqnarray}
\label{red5}
H_{\rm {AB}}=g_{2}\hat X_{\rm B}\hat Z_{\rm A}\,,
\end{eqnarray}
where $\hat Z_{\rm A}$ is an operator in the Hilbert space of $\rm A$ and
$g_{2}$ is the coupling strength. In Eq.(\ref{red5}), $\hat X_{\rm B}$
can be replaced by any operator that commutes with $\hat X_{\rm B}$.
As long as $H_{\rm {AB}}$ is of the form described by Eq.(\ref{red5}),
the state of the system $\rm S$ will evolve only to states of the form
\begin{eqnarray}
\label{red6}
\rho ^{\prime}_{\rm {AB}}= \sum_{i}|c_{i}|^2 |\phi ^{\prime }_{i}
\rangle _{\rm A} \langle \phi ^{\prime }_{i} |_{\rm A} \otimes
|X_{i}\rangle _{\rm B} \langle X_{i}|_{\rm B} \,,
\end{eqnarray}
where $|\phi ^{\prime }_{i}\rangle _{\rm A} = e^{-i g_{2}\hat X_{i} \hat
Z_{\rm A}t} |\phi_{i}\rangle _{\rm A}$ ( in which $t$ denotes the time ). 
Thus Eq.(\ref{red4}) continues to
be satisfied, and the entire information about the system $\rm S$ can be
learnt from its boundary $\rm B$.

In the above proof of the quantum holographic principle, the facts that
$\rm B$ was macroscopic enough to decohere, and that it was coupled
to both $\rm A$ and $\rm E$ by the same operator $\hat X_{\rm B}$, were 
important. We can give a counter example of the quatum holographic 
principle when the above conditions do not hold. Let $\rm S$ be a very simple
system comprised of two spin $\frac{1}{2}$ particles $\rm A$ and $\rm B$.
Only $\rm B$ directly interacts with another spin $\frac{1}{2}$ particle 
$\rm E$, and as such, is the boundary of $\rm S$. We take an initial state
of $\rm A$, $\rm B$ and $\rm E$ to be the following form
\begin{eqnarray}
\label{init}
(|{\uparrow}\rangle _{\rm A} |{\uparrow}\rangle _{\rm B} +
|{\downarrow}\rangle _{\rm A}
|{\downarrow}\rangle _{\rm B})\otimes |{\downarrow }\rangle _{\rm E}\,.
\end{eqnarray}
At this stage, the system $\rm S$ has a zero entropy and so does the
environment $\rm E$. Let the unitary interaction $U_{\rm {BE}}$ between
$B$ and $E$ be the following 
\begin{eqnarray}
\label{int}
|{\uparrow}\rangle _{\rm B}|{\downarrow }\rangle _{\rm E}
\rightarrow  |{\downarrow}\rangle _{\rm B}
|{\uparrow }\rangle _{\rm E}\,, \\ \nonumber 
|{\downarrow}\rangle _{\rm B} |{\downarrow }\rangle _{\rm E}
\rightarrow |{\downarrow}
\rangle _{\rm B} |{\downarrow }\rangle _{\rm E} \,.
\end{eqnarray}
Then the final state of $A$, $B$ and $E$ is as follows
\begin{eqnarray}
\label{fin}
(|{\uparrow}\rangle _{\rm A} |{\uparrow}\rangle _{\rm E}+ 
|{\downarrow}\rangle _{\rm A}
|{\downarrow}\rangle _{\rm E} ) \otimes |{\downarrow}\rangle _{\rm B}\,.
\end{eqnarray}
All entropy of the system $\rm S$ is stored in $A$. The boundary $\rm B$ 
has no entropy. This counter example illustrates the importance of the 
Hamiltonians $H_{\rm {BE}}$ and $H_{\rm {AB}}$ being of the form given by
Eqs.(\ref{red0}) and (\ref{red5}) for the quantum holographic 
principle to hold true.
%%%%%%%%%%%%%%%%%%%%%%%%%%%%%%%%%%%%%%%%%%%%%%%%%%%%%%%%%%%%%%%%%%%%%%%%
\section{CONCLUSION}
We have shown that under a specific set of conditions, the entire 
information about a quantum system $\rm S$ can be obtained from it's
subsystem $\rm B$ which directly interacts with the environment. In the
language of quantum measurement theory, the boundary $\rm B$ acts as
an apparatus for the state of the whole system $\rm S$. While the 
standard (semi-classical) holographic principle helps in making 
the $3$ dimensional world effectively $2$ dimensional, the 
quantum version could lead to a reduction of the Hilbert space dimensions
of a problem.

For the quantum version of the principle to reduce to the semi-classical
version, we require:

$(1)$ The physical boundary of the system in the semi-classical version 
to coincide with the subsystem $\rm B$. In other words, we require the
boundary to be interacting strongly with the enviornment and the bulk
to have insignificant interaction with the outside world.

$(2)$ The boundary itself to have quite a large number of degrees of
freedom, so that it is macroscopic and strongly decohering (i.e nearly
a classical system). It must also be coupled both to it's enviornment
and to it's bulk by the same operator $\hat X_{\rm B}$.

$(3)$ The system to be an open system. So it would not apply to closed
systems such as the entire universe.

To conclude, we emphasize that there may be other methods to derive the
holographic principle which apply to less restricted 
situations. But if we assume that all systems are essentially quantum,
and gain entropy only from interaction with their environment, then 
the requirements $(1)- (3)$ are probably essential for the 
validity of the standard
holographic principle.
%%%%%%%%%%%%%%%%%%%%%%%%%%%%%%%%%%%%%%%%%%%%%%%%%%%%%%%%%%%%%%%%%%%%%%%%%
\acknowledgments
Authors are supported by the INLAKS foundation and the ORS award. 
%%%%%%%%%%%%%%%%%%%%%%%%%%%%%%%%%%%%%%%%%%%%%%%%%%%%%%%%%%%%%%%%%%%%%%%%%


\begin{references}

\bibitem{thooft} G. 't Hooft, {\it Dimensional reduction in quantum gravity},
      gr-qc/9310026; L. Susskind, J. Math. Phys. {\bf 36}, 6377 (1995).

\bibitem{bekenstein} J. D. Bekenstein, Phys. Rev. {\bf D 23}, 287 (1981).

\bibitem{maldacena} J. Maldacena, Adv. Theor. Math. Phys. {\bf 2}, 231
         (1998), hep-th/9711200.
\bibitem{zurek91} W. H. Zurek, Phys. Today {\bf 44}, 36 (1991).

\bibitem{zurek81} W. H. Zurek, Phys. Rev. {\bf D 24}, 1519 (1981).

\bibitem{zurek} W. H. Zurek, S. Habib, and J. P. Paz, Phys. Rev. Lett
         {\bf 70}, 1187 (1993).

\bibitem{schu} B. Schumacher, Phys. Rev. {\bf A 51}, 2738 (1995).
\end{references}
\end{document}